# STELLAR-$k_{off}$: A Transfer Learning Model for Protein-Ligand Dissociation Rate Constant Prediction Based on Interaction Landscape

Jingyuan Li[1], Renxiao Wang[1,*]

[1] Department of Medicinal Chemistry, School of Pharmacy, Fudan University, Shanghai, 201203

## Abstract

The key to successful drug design lies in the correct comprehension of protein-ligand interactions. Within the current knowledge paradigm, these interactions can be described from both thermodynamic and kinetic perspectives. In recent years, many deep learning models have emerged for predicting the thermodynamic properties of protein-ligand interactions. However, there is currently no mature model for predicting kinetic properties, primarily due to the scarcity of kinetic data. To tackle this problem, we have developed a graph neural network model called STELLAR-$k_{off}$ (Structure-based TransfEr Learning for Ligand Activity Regression) to predict protein-ligand dissociation rate constant. Unlike conventional structure-based protein-ligand property prediction models, STELLAR-$k_{off}$ does not use a single structure of the protein-ligand complex as input. Instead, it employs a set of ligand conformations within the protein pocket. First, a set of ligand conformations in the protein pocket is generated through molecular docking. Then, using transfer learning, each conformation is transformed into protein-ligand interaction features via a protein-ligand affinity prediction model. Each conformation is treated as a node in a graph structure, constructing a protein-ligand interaction landscape. Finally, an equivariant graph neural network is used to predict dissociation rate constants. In addition, we expanded the PDBbind $k_{off}$ dataset from 680 to 1062 entries and employed the augmented dataset for model training and testing. When tested through five-fold cross-validation, our model exhibited outstanding performance on both validation set and external set. We believe this study provides an effective tool for predicting protein-ligand dissociation rate constant and offers new insight for the future development of this field.

# 1   Introduction

Combination of drugs and target proteins is the key to the pharmacological effect of drugs. In early stages of drug discovery, thermodynamic properties such as half inhibition rate ($IC_{50}$), dissociation constant ($K_D$), and inhibition constant ($K_i$) were measured to evaluate the binding affinity between candidate compounds and target proteins. However, protein-ligand combination occurs within a dynamic system that continuously interacts with the environment, kinetic properties can provide more detailed information about protein-ligand interactions.[1] As shown by the formula $K_D=k_{off}/k_{on}$, complexes with the same affinity may have different binding and unbinding rates, which may result in significant changes in the efficacy of drug.[2] Therefore, in the process of drug development, we should pay attention to both the kinetic and thermodynamic properties of the protein-ligand complex.[3]

Protein-ligand dissociation rate constant ($k_{off}$) is one of the kinetic properties which directly reflects the residence time of a ligand on its target protein and is highly correlated with the pharmacokinetics of drug in vivo.[4–7] The primary experimental methods for measuring $k_{off}$ include surface plasmon resonance (SPR), radioligand binding and time-resolved fluorescence resonance energy transfer. However, these techniques are time-consuming and susceptible to measurement errors due to variations in experimental conditions. Kinetics-driven drug design approaches require new technologies to enhance the accuracy and reproducibility of kinetic property assessments.

Over the past decade, molecular dynamics (MD) simulation has become the main computational method for predicting protein-ligand kinetic properties.[8] These emerging computational approaches combine with experimental methods, have significantly advanced the study of protein-ligand kinetics.[9–12] However, the MD method is difficult to be applied in high-throughput drug screening because of its cumbersome operation and large computational resource requirements. It is worth noting that although artificial intelligence (AI) approaches have developed rapidly in drug discovery, most of the approaches developed revolve around thermodynamic properties.[13–15] In the prediction of kinetic properties, the application of AI in predicting protein-ligand kinetic properties remains considerably limited, with two primary focuses: first, as an auxiliary tool in molecular dynamics simulations to assist in predicting essential simulation parameters;[12,16] and second, for directly predicting kinetic properties of specific protein targets.[17,18] However, these approaches are not yet feasible for high-throughput screening applications. Compared to the researches on protein-ligand thermodynamic properties, researches on kinetic properties commenced later, leading to a scarcity of available data. For instance, the PDBbind database contains 27,408 entries for protein-ligand affinity data, whereas data on dissociation rate constants is limited to only 680 entries. This scarcity poses a significant challenge for deep learning models, which are highly data-dependent, often resulting in severe overfitting during training and consequently limiting the applicability of AI-based approaches in predicting protein-ligand kinetic properties.[19–21]

To advance the AI-based method of protein-ligand kinetic property prediction, we expanded basic dataset and developed a deep learning model called STELLAR-$k_{off}$ (Structure-based TransfEr Learning for Ligand Activity Regression) to predict protein-ligand dissociation rate constants. In terms of dataset amplification, we increased size of the PDBbind $k_{off}$ dataset[19] by collecting new data and merging existing datasets, this expanded dataset now includes 1062 data from 222 protein targets. Regarding the model, it has long been established that the protein-ligand energy landscape is strongly correlated with dissociation rate constant,[22,23] thus we used molecular docking to generate a distribution of ligand conformations within protein pocket and utilized a pre-trained protein-ligand affinity prediction model to transform various conformations of complex into protein-ligand interaction feature matrices. This interaction landscape will be used alongside the global protein features graph as input for the equivariant graph neural network to compute the dissociation rate constants. Finally, we conducted five-fold cross-validation to evaluate the performance of our model and established two external tests involving protein complexes that were not included in the training set to assess the generalization ability. Additionally, we designed feature ablation experiments to demonstrate the effectiveness of the interaction landscape and the transfer learning approach in our model.

## 2 Method

### 2.1 Expanding PDBbind $k_{off}$ Dataset

PDBbind $k_{off}$ dataset is one of the largest publicly available datasets of protein-ligand dissociation rate constants, contains $k_{off}$ data from all PDBbind database[24] accumulated references before 2020, totaling 680 complex structures. To address the challenges posed by limited data for training deep learning models, this study expanded the PDBbind $k_{off}$ dataset. New data comes from two sources:

(1) New references in PDBbind v2021. During 2021, PDBbind dataset has accumulated a total of 3222 new references, we firstly employed a keyword-based filtering program to identify 523 references related to dissociation rate constants. Then, these references were manually read to extract complexes with curate $k_{off}$ value, we also recorded some key information about these complexes such as experimental method. Finally, the extracted complex data were cleaned using the same standards as PDBbind $k_{off}$ v2020, resulting in 59 complex entries.

(2) Non-overlapping data from other $k_{off}$ dataset. Considering the possibility that $k_{off}$ data might exist in references not accumulated in PDBbind, we surveyed other $k_{off}$ datasets published before 2020. Finally, we selected dataset complied by Nurlybek Amangeldiuly et al.[20], which comprises 501 complex entries and show minimal overlap with our dataset. The overlapping data in this dataset has been identified and removed using a combination of manual inspection and ligand SMILES comparison. We also manually read the references for each data entry and applied the same criteria as PDBbind $k_{off}$ v2020 to filter the data, resulting in 323 complex entries.

Through the above methods, we collected a total of 382 new data entries. However, 38 of these entries lack corresponding protein-ligand complex crystal structures. To provide high-quality structural data, we constructed protein-ligand complex models for those entries lacking corresponding crystal structures, then we optimized all complex structures using the same data processing process as PDBbind $k_{off}$ v2020.[19] For the entries with complex crystal structure, we used 2-nanosecond constrained molecular dynamics simulation to optimize; For those entries without crystal structure, we first construct the complex structure by molecular docking with Schrodinger software (Schrödinger LLC, version 2019),[25] and then optimize it by 2-nanosecond unrestricted molecular dynamics simulation. All the short-length MD simulations we used in this study were supported by AMBER software (version 2022).[26]

### 2.2 Model Architecture

*Generation of Ligand Conformational Distributions*

To obtain protein-ligand interaction landscapes, we must sample the conformational distribution of ligand within protein pocket. The semi-flexible molecular docking of AutoDock Vina v1.2.5[27] was used to generate as many conformations of the ligand as possible within protein pocket, forming a low-energy conformation ensemble. After testing, the optimal parameters for docking are as follows:

(1) To ensure that ligand can fully explore entire protein binding pocket, the docking box size was set to 25Å (size=25Å), and a maximum of 250 conformations were generated for each complex (num_modes=250).

(2) To achieve a more diverse conformational sampling of the ligand within the pocket, the Vina scoring range for generating conformations was set to 20 (energy_range=20).

(3) To accelerate the speed of conformation generation, the sampling exhaustiveness was set to 16 (exhaustiveness=16).

*Protein-Ligand interaction landscape Encoding and Extraction Module.*

The Protein-Ligand interaction landscape is represented as an implicit graph, where each node denotes one conformation. The feature extraction model is consisted of the equivariant graph neural network (EGNN) designed by Satorras et al.[28] EGNN is particularly suited for our task because it has proven to be translationally, rotationally, and reflectionally equivariant and computationally more efficient than other models. In addition, we adapted a well-trained protein-ligand affinity prediction model GIGN designed by Calvin Yu-Chian Chen et al.[29] into a transfer learning model to encode the protein-ligand interactions. The GIGN model is a deep learning model that takes the 3D structure of protein-ligand complex as input to predict protein-ligand affinity. After training the model using PDBbind v2020 dataset, it achieved a Pearson correlation of 0.827 on CASF2016[30] test set, indicating that GIGN model effectively captures protein-ligand interactions. Given a set of conformations for a complex, each conformation with protein structure will be fed in to the GIGN model. Then the output from graph network layer of GIGN will be extracted as node feature $\{r_i^{ini} \in \mathbb{R}^{256}\}_{i=1}^{N_r}$ for corresponding conformation and the location for each node $\{x_i \in \mathbb{R}^3\}_{i=1}^{N_r}$ is the coordinate of conformation mass center. The edge value between two nodes is the relative RMSD of corresponding conformations which is calculated by AutoRMSD (http://www.sioc-ccbg.ac.cn/software/AutoRMSD/) $\left\{a_{ij} = S_{similarity_{ij}} = 1 - \frac{RMSD_{ij}}{\max_{RMSD}}\right\}_{i=1, j=1}^{N_r}$. $N_r$ denotes the number of conformations. Then, the initial features are transformed through an embedding layer (eq1).

$$r_i^0 = f(w^{ini} r_i^{ini})$$
$$f(x) = \max(0, x) + 0.1 \min(0, x)$$
(eq1)

Where $f$ represents a LeakyReLU activation function. $w^{ini}$ represents a single linear layer with learnable parameters that transforms the input features to a latent space of 128 dimensions. Hereafter, $w$ shares the same definition, if not specified. All nodes were updated through EGNN which has edge value for several iterations (eq2).

$$m_{ij} = \phi_e\left(r_i^l, r_j^l, \|x_i^l - x_j^l\|^2, a_{ij}\right)$$
$$m_i = \sum_{j \neq i} m_{ij} \quad \text{(eq2)}$$
$$r_i^{l+1} = \phi_h(r_i^l, m_i)$$

Where $\phi_e$ denotes two consecutive linear layers with two SiLU activation functions and $\phi_h$ denotes two consecutive linear layers with one SiLU activation functions, $[,]$ represents the concatenation of vectors. After two EGNN update iterations, the features of all conformation $(r_i^{con})$ are obtained.

*Global Protein Pocket Feature Encoding and Extraction Module*

The protein pocket, defined by amino acid residues surrounding each ligand conformation, is represented as an implicit graph where each node corresponds to a pocket residue. Protein residues within a 5Å distance of each ligand conformation will be considered as neighboring residues. By taking the union of the neighboring residues from all conformations, we can obtain the complete set of neighboring residues for the global protein pocket. The initial feature for each node $\{v_p^{ini} \in \mathbb{R}^{20}\}_{p=1}^{N_v}$ is encoded by BLOSUM62 matrix[31] and the location for each node $\{x_p \in \mathbb{R}^3\}_{p=1}^{N_v}$ is the coordinate of the residue mass center. $N_v$ denotes the number of pocket residues. An embedding layer is similarly employed to transform initial features to a vector of length 32 (eq1). All nodes were updated through EGNN without edge value for several iterations (eq3). After two EGNN update iterations, the features of protein pocket $(v_p^{prot})$ are obtained.

$$m_{pq} = \phi_e(v_p^l, v_q^l, \|x_p^l - x_q^l\|^2)$$
$$m_p = \sum_{q \neq p} m_{pq} \quad \text{(eq3)}$$
$$v_p^{l+1} = \phi_h(v_p^l, m_p)$$

*Feature Aggregation Module*

In the previous module, the protein-ligand interaction landscape and global protein feature graph were updated using two individual EGNN networks. To merge these features, we designed a unidirectional cross-attention network. This network computes an attention matrix for each pocket residue with respect to each ligand conformation and concatenates the attention matrix at the end of the updated protein-ligand interaction landscape (eq4).

$$Q_r = w_r^{query} r_i^{con}$$
$$K_v = w_v^{key} v_j^{prot}$$
$$V_v = w_v^{value} v_j^{prot}$$
$$\alpha_{ij} = \text{softmax}\left(\frac{Q_r K_v^T}{\sqrt{N_v}}\right) \quad \text{(eq4)}$$
$$v_j^{update} = \alpha_{ij} V_v$$
$$r_i^{agg} = w_i^{agg}\left(r_i^{con}, v_j^{update}\right) + b_i^{agg}$$

Where T denotes matrix transpose, $b_i^{agg}$ denotes the bias of linear layer. $K_v$, $V_v$ and $Q_r$ are the key, value, query of corresponding features. The features after merging $r_i^{agg}$ are vectors of length 128.

*Dissociation Rate Constant Prediction Module*

We used global average pool to encapsulate the node-level representations after the message aggregation module into a graph-level vector as (eq5)

$$r_{ave} = \frac{1}{N_r} \sum_{i=1}^{N_r} r_i^{agg} \quad \text{(eq5)}$$

An MLP which contains two fully connected layers is used to fit $r_{ave}$ to the pk$_{off}$. The loss function used during model training is MSE (eq6).

$$Loss_{mse} = \frac{1}{N} \sum_{i=1}^{N} (y_i - \hat{y}_i)^2 \quad \text{(eq6)}$$

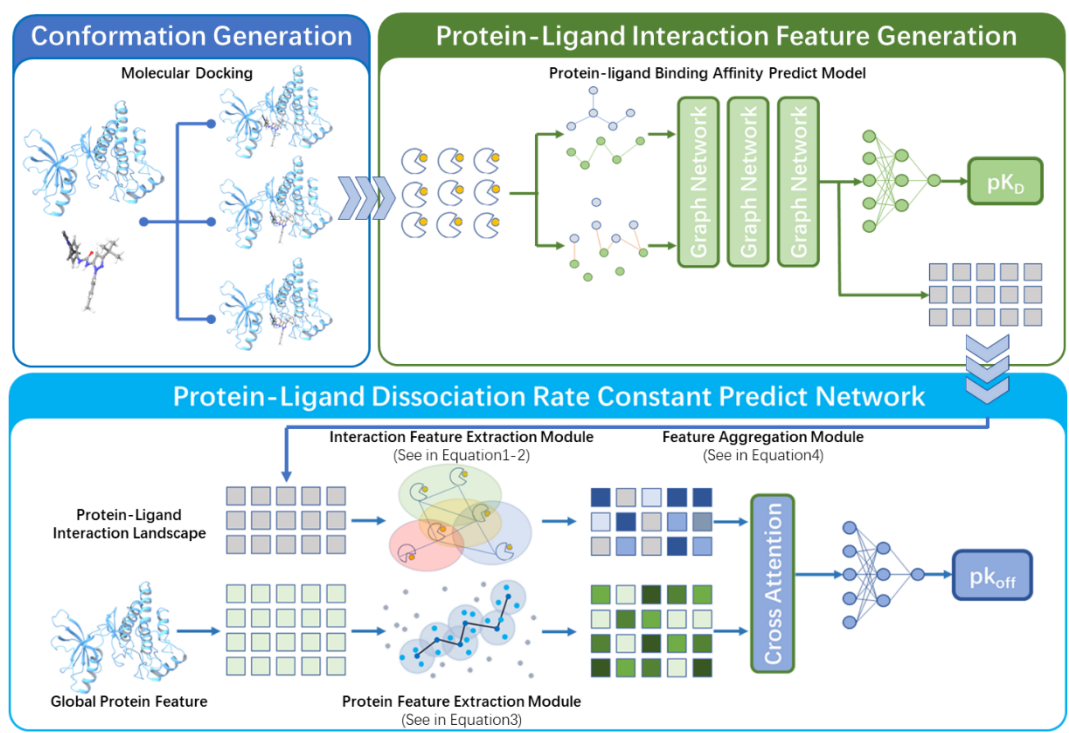

**Figure 1.** Illustration of the basic framework of STELLAR-$k_{off}$. STELLAR-$k_{off}$ firstly employs molecular docking to generate conformation ensemble. These conformations are transferred to interaction landscape and global protein graph by a protein-ligand affinity prediction model. Then STELLAR-$k_{off}$ implements modules that extract features from two graphs and merges features by cross-attention layer. Finally, the fully connected layers is used to fit features to the $pk_{off}$.

### 2.3 Model Training

*Structure Preparation*

All the data used for training were sourced from the initial structure set of the new PDBbind $k_{off}$ dataset, a total of 1,062 protein-ligand complexes. For each protein-ligand complex in this dataset, structure preparation was performed using pymol software[32] to remove water and ligand hydrogen atoms. The result protein structure was saved in the PDB format, and the ligand structure was saved in mol2 format. Subsequently, conformation ensembles for all complexes will be generated through molecular docking, several complexes could not generate more than 125 conformations were removed. As resulted, a total of 920 protein-ligand complexes were processed through the above procedure.

*The Training Process*

During model training, a batch size of 16 samples was used and the training process was performed with an Adam optimizer with an initial learning rate of 0.0001 for updating parameters. The model parameters were saved each time the RMSE on validation set decreased. Training was terminated if the RMSE on validation set did not decrease for consecutive 1000 iterations and the last saved parameters were used as the final model parameters. The whole training process was completed in 6 hours on a single NVIDIA GeForce RTX 4090 card with 24 GB memory.

## 2.4 Evaluation of the Models

*Five-fold Cross-Validation Study*

Five-fold cross-validation is a method used to evaluate the performance of machine learning models. This method allows for comprehensive utilization of the data, ensuring that each sample participates in both training and validation, thereby this method is employed to evaluate STELLAR-$k_{off}$. We extracted the complexes of p38 MAP kinase (p38) and focal adhesion kinase (FAK) from the dataset and employed as two external test sets to evaluate the generalization ability of our model. The remaining complexes were randomly divided into five equal-sized subsets by k-fold algorithm.[33] Then, the process involves five rounds of training and validation. In each round, one subset is selected as the validation set, while the remaining four subsets are used for training. After training the model in each round, its performance is evaluated on the validation set and external test sets. Finally, the average of the results from all five rounds is calculated to represent the overall performance of the model. Additionally, we designed three models to further validate the effectiveness of our model. The first model is a baseline model that takes the SMILES representation of the ligand as input, with the entire model constructed using a fully connected network. The second model involves replacing the transfer learning descriptors with RF-Score descriptors[34] to represent the protein-ligand interactions of each conformation, the structure of this model is identical to STELLAR-$k_{off}$. The last involves freezing the graph network layer parameters of the GIGN model, while fine-tuning the remaining parameters using $k_{off}$ data. For all the models above, we used Pearson correlation coefficient $(r_p)$ and the Root mean squared error $(RMSE)$ to evaluate the predictions. Below, we provide formulas to describe these metrics.

$$r_p = \frac{\sum_{i=1}^{n}(x_i - \bar{x})(y_i - \bar{y})}{\sqrt{\sum_{i=1}^{n}(x_i - \bar{x})^2 \sum_{i=1}^{n}(y_i - \bar{y})^2}} \quad (eq7)$$

$$RMSE = \sqrt{\frac{1}{n}\sum_{i=0}^{n-1}(y_i - \hat{y}_i)^2} \quad (eq8)$$

where $\hat{y}_i$ is the predicted value of the ith sample, $y_i$ is the corresponding ground truth, n is the number of samples, and $\bar{y}$ is the mean value of the vector y, analogously to $\bar{x}$.

*Ablation Study*

To evaluate the impact of protein-ligand interaction landscape, global protein pocket feature, and conformation similarities on the prediction of protein-ligand dissociation rate constants in STELLAR-$k_{off}$, we designed an ablation experiment. Specifically, we retrained the model after removing each feature individually while keeping the remaining features. By comparing the performance of the full model with that of the models trained with specific features removed on the same test set, we quantified the contribution of each feature to our model.

# 3 Result and Dissuasion

A powerful deep learning model requires not only a well-designed architecture but also ample high-quality training data. To train a deep learning model capable of accurately predicting $k_{off}$, more high-quality data is essential. In previous work, the PDBbind $k_{off}$ v2020 dataset collected $k_{off}$ data from references accumulated by the PDBbind database until 2020. However, the PDBbind database primarily focuses on references related to binding affinity, collecting data solely on these references is insufficient. We collected new records from two sources, new references in PDBbind v2021 and non-overlapping data in other datasets established before 2020. Most of the new records have the corresponding complex structures resolved by X-ray crystal, and we download the structures from the Protein Data Bank (PDB)[35]. For the remaining ones, we used molecular docking and short-time MD to build complex structures. Finally, we gathered 382 new entries in total and the total number of entries has increased to 1062. Below, a distribution map of $pk_{off}$ values is shown (Figure 2a), the new dataset contains a more diverse and abundant range of data compared to the previous version. In addition, the entire data set was further clustered by the sequence of protein in each complex using the CD-hit program (v4.8.1)[36], with the similarity threshold setting to 90%. The clustering results indicated that all samples in the data set could be divided into 222 groups. Comparing to previous dataset, the number of protein groups has increased by 30% (Figure 2b). Overall, this data augmentation is highly significant, both in terms of the distribution of dissociation rate constants and the diversity of protein targets, the new dataset shows a marked improvement in data diversity. Compared with existing biochemical kinetics databases, as shown in Table 1, our dataset contains significantly more entries than most others, second only to the non-public Pfizer Database.[37] This dataset, including the experimental $k_{off}$ values the 3D structures of all complexes and some other information s, is free available from our new PDBbind-CN web site (https://www.pdbbind-plus.org.cn/).

**Table 1.** Databases of Biomolecular Binding Kinetics

| Database | Descriptor | Website | Stats |
|---|---|---|---|
| PDBbind-koff v2021 | A dataset of protein-ligand dissociation rate constants, containing 1,061 protein-small molecule dissociation rate constants with corresponding complex structures, and providing short-term dynamic simulation results of the complexes. | https://www.pdbbind-plus.org.cn/ | Maintained |
| Nurlybek Amangeldiuly's Dataset[20] | A dataset of protein-ligand dissociation rate constants, containing 501 entries with corresponding complex structures. | None | Unmaintained, the data can be downloaded from the reference |
| BindingDB[38] | A dataset of protein-ligand dissociation rate constants, containing 301 entries with corresponding complex structures. | https://bindingdb.org/rwd/bind/ByKI.jsp?specified=Kn | Unmaintained. |
| KOFFI[39] | A biochemical kinetics dataset containing 1,705 kinetic data entries, primarily consisting of protein-protein and protein-nucleic acid kinetic data, with fewer entries for protein-small molecule interactions. | http://koffidb.org/ | Unmaintained. |
| Pfizer Dataset[37] | A protein-ligand dynamics dataset containing 2,046 dissociation rate constant entries, with structural details unspecified. | None | Not publicly available |

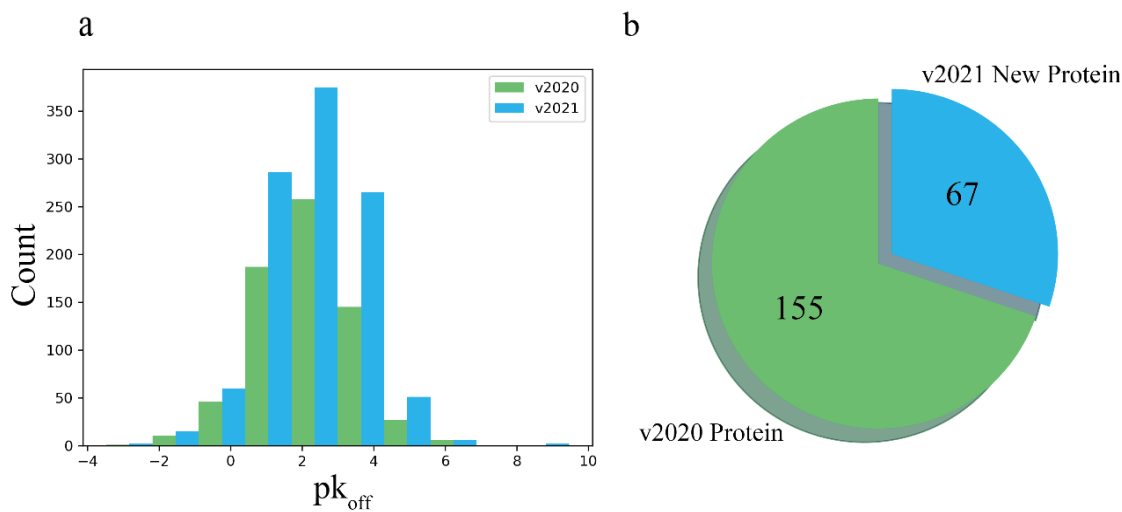

**Figure 2.** Statistical result of PDBbind $k_{off}$ Dataset v2021. (a) Distribution of the $pk_{off}$ values for 1062 complexes in new dataset. (b) Protein cluster result of new dataset.

For STELLAR-$k_{off}$ model, we employed all the data in PDBbind $k_{off}$ Dataset v2021 for training and testing our model. To accurately evaluate the performance of our model, we used the five-fold cross-validation method which is mentioned in the methods section, and the conclusion were obtained by averaging each cross-validation result. The correlation between the experimental dissociation rate constants and the values predicted by STELLAR-$k_{off}$ on validation set is shown in Figure 3, where the averaged Pearson correlation coefficient is 0.729 and Root mean squared error is 0.905. Moreover, our model produced a Pearson correlation coefficient of 0.838 on focal adhesion kinase set and 0.697 on p38 MAP kinase set, this indicates that our model exhibits strong generalization ability, as it can still provide reasonably accurate predictions even when the protein structures in the protein-ligand complexes given to the model are not part of the training set. The basic information on other machine learning models or molecular dynamic simulation methods published in recent years is summarized in Table 1. Our model demonstrates strong performance compared to the RF model developed by Su et al.[19], especially in predicting the $k_{off}$ of complexes on focal adhesion kinase, this suggests that our model not only demonstrates strong prediction accuracy but also exhibits a notable degree of generalization capability.

However, the random forest model developed by Nurlybek Amangeldiuly et al.[20] seems to surpass our model in performance (Average $r_p$=0.78), this may because machine learning models generally have lower data requirements compared to deep learning models, also it is important to note that we have more test data than theirs, the increase in both data quantity and diversity will elevate the complexity of the task. STELLAR-$k_{off}$ also demonstrates comparable performance to several kinetic simulation methods, but STELLAR-$k_{off}$ as a deep learning model, it offers significantly superior computational speed. STELLAR-$k_{off}$ can predict dissociation rate constants for over 400 complexes within a single day, requiring only the independent 3D structures of the protein and ligand, along with pocket center coordinates. This approach eliminates the need for carefully constructed protein-ligand complex structures, significantly enhancing usability and accessibility for researchers. Thus, we believe that with the continuous expansion of datasets in the future, deep learning will emerge as a more promising approach.

**Table 2.** Performance of STELLAR-$k_{off}$ and Other Baseline Methods for Protein-Ligand Dissociation Constant Prediction

|  |  | Validation set | FAK set | p38 set |
|---|---|---|---|---|
| STELLAR-$k_{off}$ | Average $r_p$ | 0.729 (N=846) | 0.838 (N=33) | 0.697 (N=41) |
|  | Average RMSE | 0.905 (N=846) | 0.625 (N=33) | 0.997 (N=41) |
| Nurlybek Amangeldiuly's RF model[20] | Average $r_p$ | 0.78 (N=501) | NA | 0.75 (N=28) |
|  | Average RMSE | 0.82 (N=501) | NA | 1.10 (N=28) |
| Su's RF model[19] | $r_p$ | 0.706 (N=102) | 0.241 (N=33) | NA |
|  | RMSE | 0.986 (N=102) | 1.044 (N=33) | NA |
| Wong's SMD[40] | $r_p$ | NA | 0.76 (N=14) | NA |
| | RMSE | NA | NA | NA |
| Wade's COMBINE[41] | Average $r_p$ | NA | NA | NA |
| | Average RMSE | NA | NA | 0.88 (N=22) |

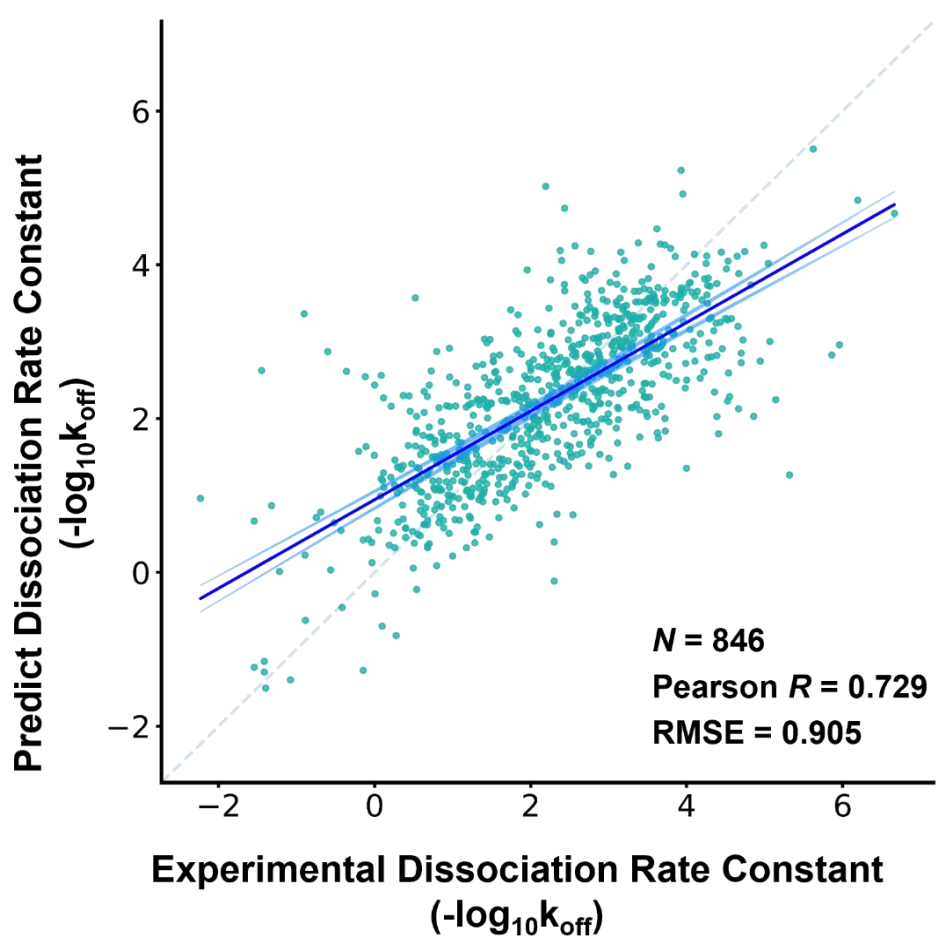

**Figure 3.** Correlation between the experimental dissociation rate constants (p$k_{off}$) and the predicted values in five-fold crossing-validation (N = 846; Rp = 0.729; RMSE = 0.905 log units). The solid line is the regression, where 95% confidence interval is indicated by the shaded region.

To evaluate the importance of the conformation ensemble and transfer learning descriptors to STELLAR-$k_{off}$, we designed the baseline model based on ligand SMILES, the GIGN fine-tuned model and the STELLAR-$k_{off}$ model based on RF-Score descriptors for comparison. As shown in Table 2, the original STELLAR-$k_{off}$ model outperforms the other three models across all the test sets. In the test of GIGN fine-tuned model, the GIGN model uses the protein-ligand bound state as input, predicting affinity based on protein-ligand interactions. However, protein-ligand dissociation is a dynamic process, and the energy changes throughout the process are not only related to the bound-state conformation. This may explain the phenomenon that the GIGN fine-tuned model performs poorly in predicting $k_{off}$. In contrast, our model takes the ensemble of multiple ligand conformations within the protein pocket as input, potentially capturing conformations that significantly contribute to the energy changes during dissociation, allowing for more accurate $k_{off}$ predictions. Similarly, in the test of RF-Score based STELLAR-$k_{off}$ model, RF-Score is a primitive protein-ligand description method, it characterizes a protein−ligand complex binding site as a one-dimensional vector containing occurrence counts for protein atom−ligand atom pairs within a given radius. In the dissociation rate constant prediction task where data is scarce, this primitive description method requires more parameters to fit it into the value of $pk_{off}$, which may lead to serious overfitting of the deep learning model. By comparison, our model employed a protein-ligand affinity prediction model trained with sufficient data to generate descriptors. Consequently, the interaction features generated by our model are more sophisticated and less susceptible to overfitting, resulting in improved overall model performance. In summary, the descriptor method we designed can enhance the performance of deep learning model in predicting dissociation rate constants.

**Table 3.** Performance of Different Model Architectures Similar to STELLAR-$k_{off}$ in Predicting $pk_{off}$

|  |  | Validation set | FAK set | p38 set |
|---|---|---|---|---|
| STELLAR-$k_{off}$ | Average $r_p$ | 0.729 | 0.838 | 0.697 |
| Org | Average RMSE | 0.905 | 0.625 | 0.997 |
| Baseline Model | Average $r_p$ | 0.441 | 0.463 | 0.189 |
| | Average RMSE | 1.326 | 0.924 | 1.402 |
| GIGN Fine-tuned Model | Average $r_p$ | 0.601 | 0.542 | 0.450 |
| | Average RMSE | 1.054 | 0.975 | 1.216 |
| STELLAR-$k_{off}$ RF-Score | Average $r_p$ | 0.692 | 0.751 | 0.601 |
| | Average RMSE | 0.967 | 0.771 | 1.100 |

We developed three variants of STELLAR-$k_{off}$ to explore the impact of each component on its performance. As illustrated in Fig 4., the protein-ligand interaction landscape emerged as the most critical input feature. Omitting this feature and relying solely on the protein global feature led to a notable reduction in model performance, with the Pearson correlation coefficient in five-fold cross-validation decreasing from 0.729 to 0.667. The most substantial performance decline occurred in the external set for p38 MAP kinase, where the Pearson correlation coefficient dropped to just 0.107. Introducing the relative RMSD between ligand conformations as an edge feature slightly improved the model's performance, with the average Pearson correlation coefficient in five-fold cross-validation increasing from 0.714 to 0.729. This improvement can be attributed to the RMSD provides a more detailed representation of the spatial relationships between ligand conformations. The global protein features had the least impact on model performance, but the five-fold cross-validation revealed that incorporating these features reduced the sensitivity of STELLAR-$k_{off}$ to the training data, thereby enhancing the stability of our model. Overall, these results suggest that all three components had a certain degree of impact on performance.

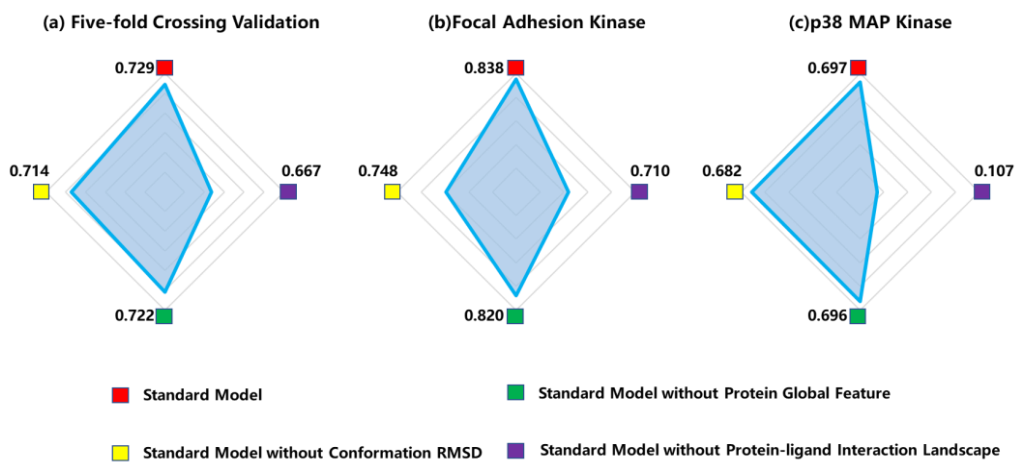

**Figure 4.** Ablation Study and Feature Importance Analysis Results. (a) Average Pearson *R* in five-fold cross-validation. (b) Pearson *R* on focal adhesion kinase. (c) Pearson *R* on p38 MAP kinase.

## 4  Conclusion

In this work, we aimed to develop a structure-based deep learning model for predicting protein-ligand dissociation rate constants. The most challenging aspect of this task is the scarcity of available data. To address this, we expanded the PDBbind $k_{off}$ dataset by collecting data from new references and merging non-overlapping datasets, increasing the total number of entries to 1,062. Based on this data, we designed and trained STELLAR-$k_{off}$ (i.e, Structure-based TransfEr Learning for Ligand Activity Regression), a transfer learning model that takes the 3D structures of proteins and ligands as input to predict dissociation rate constants. The essential feature of STELLAR-$k_{off}$ is that it utilizes a protein-ligand interaction landscape as input, rather than focusing solely on the binding conformation as other structure-based model. We first generated a set of ligand conformations in the protein binding pocket using molecular docking, followed by using a protein-ligand affinity prediction model to convert these conformations into corresponding interaction features. Finally, these features, along with the spatial relationships between conformations, were assembled into a protein-ligand interaction landscape. We believe that using these features as input allows the model to capture the overall interaction patterns within the pocket, thereby improving both the interpretability and accuracy of dissociation rate constant predictions.

Evaluations conducted on the five-fold cross-validation showed that STELLAR-$k_{off}$ demonstrated strong performance compared to existing prediction methods. Besides, STELLAR-$k_{off}$ also exhibited reasonable performance on two independent external sets, indicating a certain level of generalizability. Subsequently, we designed comparative experiments demonstrated the validity of using multi-conformations and transfer learning to generate a protein-ligand interaction landscape as model input. Finally, we designed ablation studies to evaluate the importance of each model component, confirming the crucial role of the protein-ligand interaction landscape in predicting dissociation rate constants.

With its decent accuracy and efficiency, we expect STELLAR-$k_{off}$ to become a practical tool for protein-ligand dissociation rate prediction. Additionally, we believe that the feature used in our model can open new avenues for the prediction of protein-ligand kinetic properties. In the future, as the accumulation of kinetic data continues, deep learning methods are expected to perform better in this field.